\documentclass[aps,prl,superscriptaddress,showpacs,letterpaper,twocolumn]{revtex4-1}

\usepackage{float}
\usepackage{graphicx} 
\usepackage{array}    
\usepackage{amsmath}   
\usepackage{amssymb}   
\usepackage{wrapfig}  

\usepackage{color}  
\usepackage{setspace}

\newcommand \mcF{{\mathcal F}}
\newcommand{\ket}[1]{| #1 \rangle}
\newcommand{\bra}[1]{\langle #1 |}
\newcommand \bea{\begin{eqnarray}}
\newcommand \eea{\end{eqnarray}}
\newcommand \be{\begin{equation}}
\newcommand \ee{\end{equation}}
\newcommand \nn{\nonumber}

\begin{document}	


\title{Preparation of entangled and antiferromagnetic states by dissipative Rydberg pumping}
\author{ A. W. Carr and  M. Saffman}
\affiliation{Department of Physics,  1150 University Avenue,
University of Wisconsin,  Madison, Wisconsin 53706
}

 \date{\today}

\begin{abstract}
We propose and analyze an approach for preparation of high fidelity entanglement and antiferromagnetic states using Rydberg mediated interactions with dissipation. Using asymmetric Rydberg interactions the two-atom Bell singlet is a dark state of the Rydberg pumping process. Master equation simulations demonstrate Bell singlet preparation fidelity $\mcF=0.998$. Antiferromagnetic states are generated on a four spin plaquette in agreement with results found from diagonalization of the transverse field Ising Hamiltonian. 
\end{abstract}

\maketitle


Neutral atoms are providing  a new tool for studying many body physics  and quantum magnetism\cite{Bloch2008b,*Esslinger2010,*Trefzger2011,*Baranov2012}. Recent experiments have probed strongly correlated spin systems relying on short range contact interactions between cold atoms \cite{Simon2011,*Zhang2012}, or Coulomb interactions of trapped ions\cite{Friedenauer2010,*Kim2010,*Lanyon2011,*Britton2012}.  Much recent interest has focused on antiferromagnetically ordered spin states. When the spin interactions are mediated by short-range scattering it is challenging to reach the extremely low temperatures needed to observe antiferromagnetic ordering\cite{Jordens2010}. Recent progress with Rydberg excited atoms\cite{Saffman2010,*Low2012} has demonstrated strong and long-range dipolar interactions which are suitable for creating magnetic phases with long range order\cite{Weimer2010b,*Weimer2010,*Sela2011,*Lesanovsky2011,*Ji2011,*Zeller2012}.

In this letter we propose and analyze an approach to entanglement generation and spin ordering which relies on dissipative dynamics with Rydberg state mediated interactions. 
It is well known that dissipative dynamics can be used for creating  entanglement\cite{Plenio1999,*Schneider2002,*Braun2002,*Jakobczyk2002,*Basharov2002}, and more generally  for universal quantum computational tasks \cite{Diehl2008,*Verstraete2009} as has been demonstrated in recent experiments\cite{Krauter2011,Barreiro2011}. These developments have led to a high level of activity on this topic resulting in approaches to dissipative entanglement generation in a range of physical settings\cite{Martin-Cano2011,*Murch2012,*Cho2011,*Kordas2012,*DallaTorre2013,*Tan2013}.

A dissipative approach to 
 antiferromagnetic ordering using Rydberg interactions was proposed in\cite{Lee2011}. Here we consider an arrangement with spin-dependent Rydberg interactions, which allows us to prepare the two-atom spin singlet state as a dark state of the dissipative evolution. In contrast to coherent  blockade experiments\cite{Wilk2010,*Isenhower2010,*Zhang2010} which rely on minimizing dissipation in order to maximize the fidelity of the target quantum state, the present approach exploits spontaneous emission, yet can be used to prepare a maximally entangled singlet state with fidelity exceeding 0.998. This dynamics enables high fidelity entanglement at long range which will be useful for teleporting gates in a spatially extended qubit array, as well as creating strongly correlated spin systems. Remarkably the dissipative approach described here is capable of creating the same entanglement fidelity as the coherent Rydberg blockade gate\cite{XZhang2012} but with 1500 times smaller Rydberg interaction. This implies that entanglement can be extended to much longer interparticle separations which will enable efficient computation and many particle entanglement in extended qubit arrays.

\begin{figure}[!t]
\centering
\begin{minipage}[c]{8.5cm}
\centering
 \includegraphics[width=8.5cm]{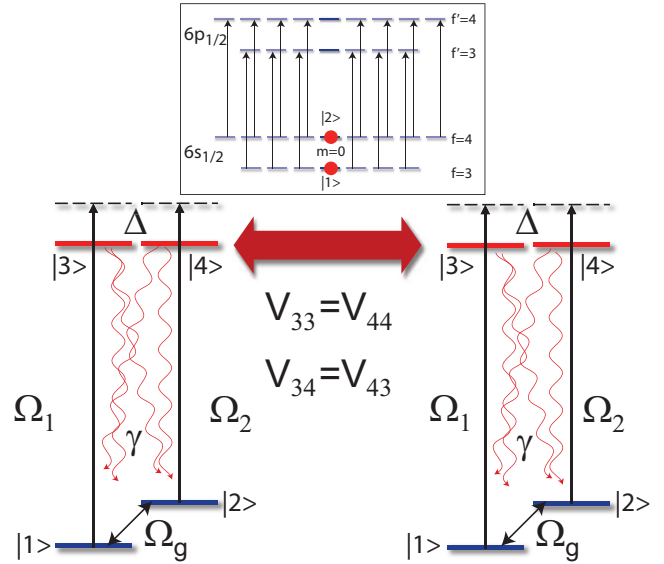}
  \caption{Level diagram for pumping into an asymmetric ground state using unequal Rydberg interactions. The boxed inset shows how $\pi$ polarized light can be used to recycle all other Zeeman ground states in the example of Cs atoms. See text for details.   }
\label{fig.afatoms}
\end{minipage}
\end{figure}

Our approach to dissipative preparation of a spin singlet is illustrated in  Fig. \ref{fig.afatoms}.
Consider two atoms with ground states $\ket{1},\ket{2}$ and Rydberg states $\ket{3},\ket{4}$. State $1$ is coupled to $3$ with Rabi frequency $\Omega_1$ and $2$ is coupled to $4$ with $\Omega_2$. The driving fields are detuned by $\Delta$. The energy splitting of the ground states $\hbar\omega_{21}$ is assumed to be a few GHz corresponding to atomic hyperfine ground states. The splitting  is  sufficiently large for us to neglect off-resonant coupling of $\ket{1}\rightarrow\ket{4}$ and $\ket{2}\rightarrow\ket{3}$.  States $3,4$ decay to $1,2$ with rate $\gamma$ and equal branching ratios. In addition there is a field coupling $\ket{1}\leftrightarrow\ket{2}$ with Rabi frequency $\Omega_g.$ This could be implemented optically with two-photon Raman transitions or with a microwave field. 

The Rydberg states interact with energy $V_{33}=V_{44}$ when both atoms are excited to the same state and energy $V_{34}=V_{43}$ when the atoms are excited to different states. In general the Rydberg interaction energy depends on angular degrees of freedom so the coupling strength will be different when $\ket{3}, \ket{4}$ correspond to different Zeeman sublevels\cite{Walker2008}. 
The idea of using asymmetric Rydberg couplings for quantum state control was previously used in several papers \cite{Brion2007c,Saffman2009b,Isenhower2011}, In those papers a very strong interaction asymmetry was required for high fidelity control. The idea described here requires only a weak asymmetry to enable a multi-atom  optical pumping process.  
We now choose a  detuning $\Delta=V_{33}/2\hbar=\Delta_{33}/2$. Thus when the ground atomic state  is $\ket{11}$ or $\ket{22}$ both atoms will be excited and spontaneously decay from the Rydberg level, thereby populating all four ground states $\ket{11},\ket{12},\ket{21},\ket{22}.$
On the other hand excitation out of states  $\ket{12}$ or $\ket{21}$
to $\ket{34}$ or $\ket{43}$  will be off-resonant by an amount $\delta= \Delta_{33}-\Delta_{34}$, and excitation of a single atom will be off-resonant by $\Delta$. 
Provided the off-resonant excitation rates are small compared to excitation of the symmetric ground states the system will be pumped into an asymmetric two-atom state.

Although spontaneous emission in real atoms populates other hyperfine ground states not in the basis $\ket{1},\ket{2}$ this can be dealt with using
recycling lasers as shown in the inset of Fig. \ref{fig.afatoms}.  Explicitly for the example of Cs atoms we take $\ket{1}=\ket{f,m}=\ket{3,0}, \ket{2}=\ket{4,0}$ and add $\pi$ polarized lasers coupling $\ket{6s_{1/2},f=3}\rightarrow \ket{6p_{1/2},f=3}$,   
$\ket{6s_{1/2},f=4}\rightarrow \ket{6p_{1/2},f=4}$. These lasers do not disturb $\ket{1},\ket{2}$ but recycle all other ground states.

We can gain some insight into appropriate parameters by considering the limit of coherent Schr\"odinger evolution. Assume two, two-level atoms with ground state $\ket{1}$, Rydberg state $\ket{3}$ and Rydberg interaction 
strength $V_{33}=\hbar \Delta_{33}.$ There is resonant excitation of 
the doubly occupied Rydberg state when the laser detuning is $\Delta=\Delta_{33}/2.$ The states $\ket{13}$ and $\ket{31}$ are off-resonant by $\Delta_{33}/2$ and are only weakly excited. 
Writing the state vector as 
\bea
\ket{\psi}&=&c_{11}(t)\ket{11}+c_{13}(t)e^{-\imath \omega_a t}\ket{13}\nn\\
&+&c_{31}(t)e^{-\imath \omega_a t}\ket{31}+c_{33}(t)e^{-\imath (2\omega_a+\Delta_{33}) t}\ket{33}\nn
\eea
the Schr\"odinger equation takes the form 
\bea
\frac{d c_{11}}{dt}&=& i\frac{(\sqrt2 \Omega^*)}{2}e^{\imath \Delta t} s\nn\\
\frac{d s}{dt}&=& i\frac{(\sqrt2 \Omega)}{2}e^{-\imath \Delta t} c_{11}+ i\frac{(\sqrt2 \Omega^*)}{2}e^{\imath (\Delta-\Delta_{33}) t}  c_{33}\nn\\
\frac{d  c_{33}}{dt}&=& i\frac{(\sqrt2 \Omega)}{2}e^{-\imath (\Delta -\Delta_{33})t} s\nn.
\eea
Here $s=\frac{1}{\sqrt2}(c_{13}+c_{31})$,  $\omega_a$ is the transition frequency, and $\Delta=\omega-\omega_a$ is the laser detuning from the non-interacting atomic transition resonance at $\omega_a$.
Adiabatically eliminating the singly excited amplitude $s$  at $\Delta=\Delta_{33}/2$ we get an effective two-level system for states $\ket{11}$ and  $\ket{33}$ resonantly coupled with the Rabi frequency $\Omega_R = (\sqrt2\Omega)^2/\Delta_{33}$.  Allowing for four-levels as in Fig. \ref{fig.afatoms}
the symmetric states are resonantly coupled to Rydberg levels with $\Omega_R$ while the antisymmetric states are also coupled with $\Omega_R$, but at a detuning $\delta=\Delta_{33}-\Delta_{34}$. 

In the absence of symmetry breaking we expect the atoms to end up in a state $\ket{\psi} \sim \ket{12} + e^{\imath\varphi}\ket{21}$ with $\varphi$ an undetermined phase. In order to create coherence between $\ket{12}$ and $\ket{21}$ we add the transverse drive $\Omega_g$. The singlet state $\ket{-}=\frac{1}{\sqrt{2}}(\ket{12}-\ket{21})$ is invariant with respect to $X$ rotations,  whereas the $m=0$ triplet state $\ket{+}=\frac{1}{\sqrt{2}}(\ket{12}+\ket{21})$ couples to states with $m=\pm 1$ which are subject to repumping via the Rydberg states. We thus expect the combined action of $\Omega_1, \Omega_2, \Omega_g$ with the detunings and Rydberg couplings specified above will drive the atoms into the maximally entangled ``antiferromagnetic" state $\ket{-}$. 

\begin{figure}[!t]
\centering
\begin{minipage}[c]{8.5cm}
\centering
 \includegraphics[width=8.5cm]{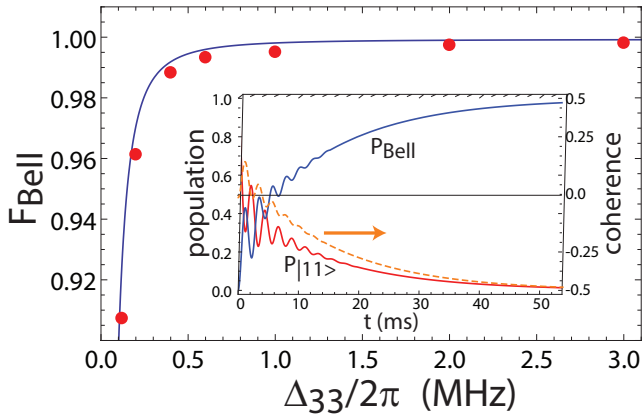}
  \caption{(color online) Comparison of $P_{\rm AF}$ from Eq. (\ref{eq.paf}) (solid line) with singlet state fidelity $F_{\rm Bell}$ from numerical solution of  Eq. (\ref{eq.master}) (red dots). Parameters: $\Omega/2\pi = 0.01 ~\rm MHz$, $\gamma=1./(0.73~{\rm ms)}$, $\Delta_{34}/\Delta_{33}=0.2$, $\Omega_R=2\Omega^2/\Delta_{33}$, and $\Omega_g= 0.5\Omega_R$. The atomic parameters correspond to the Cs $125p_{1/2}$ state with $|3\rangle,|4\rangle$ $m_j=\pm 1/2$ and the quantization axis perpendicular to the line containing the atoms. The inset shows the time dependence for $\Delta_{33}/2\pi = 1~\rm MHz$. }
\label{fig.paf}
\end{minipage}
\end{figure}

Denoting the probability for the atoms to be in the antiferromagnetic state $\ket{-}$ by $P_{\rm AF}$ the rates at which probability enters and leaves this state due to one and two-atom excitation processes are 
\begin{subequations}
\bea
r_{\rm 1,in}&=& (1-P_{\rm AF}) 2 \frac{\gamma}{4}\frac{\frac{\Omega^2}{\gamma^2}}{1+\frac{\Delta_{33}^2}{\gamma^2}+\frac{2\Omega^2}{\gamma^2}},\\
r_{\rm 1,out}&=& P_{\rm AF} 2 \frac{3\gamma}{4}\frac{\frac{\Omega^2}{\gamma^2}}{1+\frac{\Delta_{33}^2}{\gamma^2}+\frac{2\Omega^2}{\gamma^2}},\\
r_{\rm 2, in}&=&(1-P_{\rm AF})\frac{\gamma}{4}\frac{\frac{\Omega_R^2}{\gamma^2}}{1+\frac{2\Omega_R^2}{\gamma^2}},\\
r_{\rm 2, out}&=&P_{\rm AF}\frac{3\gamma}{4}\frac{\frac{\Omega_R^2}{\gamma^2}}{1+\frac{4\delta^2}{\gamma^2}+\frac{2\Omega_R^2}{\gamma^2}}.
\eea
\label{eq.rinout}
\end{subequations}
Numerical factors in these equations correspond to an idealized two-level atomic ground state. In a real atom adjustments should be made to account for branching ratios in the radiative decay from the Rydberg state. 
We then solve 
$r_{1, \rm in}+r_{2,\rm in} = r_{1, \rm out}+r_{2,\rm out}$,    to find  the equilibrium population 
\begin{widetext}
\be
P_{\rm AF}=\frac{\left(\gamma ^2+4 \delta ^2+2 \Omega_R^2\right) \left[\Omega_R^2 \left(\gamma ^2+\Delta_{33}^2\right)+2
   \Omega^2 \left(\gamma ^2+3 \Omega_R^2\right)\right]}{4 \Omega_R^2 \left(\gamma^2+\Delta_{33}^2\right)
   \left(\gamma ^2+\delta ^2+2 \Omega_R^2\right)+8 \Omega^2 \left[\gamma^4+\gamma^2 \left(4 \delta ^2+5 \Omega_R^2\right)+9 \delta ^2 \Omega_R^2+6 \Omega_R^4\right]}.
\label{eq.paf}
\ee
\end{widetext}
This expression is plotted in Fig. \ref{fig.paf} for representative parameters. As the interaction strength increases the fidelity of the Bell singlet tends to unity. This is also borne out by the limiting 
expression found by taking  $\delta\gg \gamma,\Omega_R$, 
$\Delta_{33}\gg \Omega$ so that both the one and two atom rates out of the singlet state are small, which leads to 
$P_{\rm AF} \simeq\left(1+\frac{\gamma^2}{2\Omega^2}\right)/
\left(1+\frac{2\gamma^2}{\Omega^2}\right).
$ We see that provided $\Omega\gg \gamma$ then  $P_{\rm AF}\rightarrow 1$.

To verify the approximate rate equation analysis  we have numerically solved the two-atom master equation 
\begin{equation}
\frac{d\rho}{dt}=-\frac{i}{\hbar}[\mathcal{H},\rho]+\mathcal{L},
\label{eq.master}
\end{equation}
with  
$\mathcal{H}=\mathcal{H}_{1}\otimes I_{2}+I_{1}\otimes\mathcal{H}_{2}+\mathcal{V}$, 
$\mathcal{L}=\mathcal{L}_{1}\otimes I_{2}+I_{1}\otimes\mathcal{L}_{2}$, and  $I_{\rm 1}$, $I_{\rm 2}$ 
are  $4\times4$ identity matrices.
Working in the interaction picture and making the rotating wave approximation the one atom operators  expressed in the basis  $\{|1\rangle,|2\rangle,|3\rangle,|4\rangle\}$ are
\begin{subequations}
\begin{eqnarray}
\mathcal{H}_j & = & \hbar\left(\begin{array}{cccc}
0  & \Omega_g^*/2 & \Omega_{1}^{*}/2 & 0\\
\Omega_g/2 & 0 & 0 & \Omega_2^*/2\\
\Omega_{1}/2 &0 & -\Delta & 0\\
0& \Omega_{2}/2   & 0& -\Delta\end{array}\right),\label{eq:Hamiltonian}\\
\mathcal{L}_j&=&-\frac{1}{2}\sum_{k,l=1}^4\gamma_{kl}\left( \sigma_{kl}^{(j)}\sigma_{lk}^{(j)}\rho+\rho\sigma_{kl}^{(j)}\sigma_{lk}^{(j)}-2\sigma_{lk}^{(j)}\rho\sigma_{kl}^{(j)}\right).\nonumber\\
\label{eq:Liouville}
\end{eqnarray}
\label{eq.HL1}
\end{subequations}
In the expression for $\mathcal{L}_j$, $\sigma_{kl}^{(j)}=\ket{k}^{(j)} ~^{(j)}\bra{l}$ are one-atom operators acting on atom $j$,
$\gamma_{31}=\gamma_{32}=\gamma_{41}=\gamma_{42}=\gamma/2$ and all other $\gamma_{kl}=0$. 
 The interaction term is 
\begin{eqnarray} 
\mathcal{V}&=&\hbar \Delta_{33}\, \left(\ket{33}\bra{33}+\ket{44}\bra{44} \right) + \hbar\Delta_{34}\, \left(\ket{34}\bra{34}+\ket{43}\bra{43} \right).\nonumber
\end{eqnarray}
Numerical solutions of (\ref{eq.master})  are used to extract the fidelity of the Bell singlet state $\ket{-}$ given by $F=\frac{1}{2}(\rho_{11;22}+\rho_{22;11})+|\rho_{12;21}|.$ Values of $F$ found from integrating to $t=90/\Omega_R$ starting from the initial condition 
$\ket{11}$ are compared in Fig. \ref{fig.paf} with the approximate result for $P_{\rm AF}$ that comes from solving Eqs. (\ref{eq.rinout}).

We see that the approximate result agrees well with numerical solutions. 
The maximum Bell state fidelity is $F_{\rm Bell}=0.9988$ at $\Delta_{33}/2\pi = 3~\rm MHz$. This is essentially the same fidelity as the best found in \cite{XZhang2012} for the coherent Rydberg blockade controlled phase gate. 
It is noteworthy that the dissipative approach does not require single atom addressing and the fidelity is achieved with about 1500 times smaller Rydberg interaction strength. This implies that high fidelity entanglement can be achieved at very much larger atomic separations than for the coherent interaction.  A viable approach to long range gates in an array of qubits could thus be based on teleportation\cite{Gottesman1999} using the dissipative mechanism for establishing entanglement, followed by short range coherent gates between neighboring qubits. 

 Our calculations ignore undesired entanglement between spin and center of mass degrees of freedom. This can be suppressed, despite the presence of spontaneous emission from the Rydberg levels, provided we confine the atoms in  the Lamb-Dicke regime and use magic ground-Rydberg trapping potentials\cite{SZhang2011}.

The Ising model with transverse field can be written as
\be
{\mathcal H}= J \sum_i \sigma_z^{(i)} \sigma_z^{(i+1)}+B \sum_i \sigma_x^{(i)},~~~~ J>0
\label{eq.Hising}
\ee
where $J$ is the spin-spin interaction strength and $B$ is the field strength. 
With $J>0$ the antiferromagnetic state with neighboring spins antiparallel is trivially the ground state when we restrict to nearest neighbor couplings, and there is no frustration or transverse field. For large transverse field strengths the ground state has all spins aligned along $x$. 
Finding the ground state of $\mathcal H$ on a 2D lattice with couplings that extend beyond nearest neighbors and with a local transverse field  is generally a hard computational problem\cite{Barahona1982} which may be amenable to simulation using the 
Rydberg couplings described above.

\begin{figure}[!t]
\centering
\begin{minipage}[c]{8.5cm}
\centering
 \includegraphics[width=8.5cm]{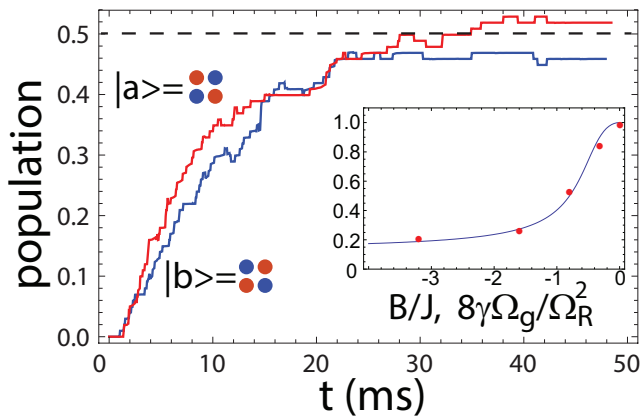}
  \caption{Population of $AF_\pm$ states from Monte-Carlo simulations of Eqs. (\ref{eq.master},\ref{eq.HL1}) on a square plaquette and averaged over 100 trajectories. The blue and red curves give the populations of $|1212\rangle$ and $|2121\rangle$. The inset shows the total population in states
$|1212\rangle$ and $|2121\rangle$ as a function of $B/J$ from numerical diagonalization of (\ref{eq.Hising}) (solid blue line) and from
 Monte-Carlo simulations of Eq. (\ref{eq.master}) (red dots). Numerical parameters were  $\Omega/2\pi = 0.01 ~\rm MHz$, $\gamma=1./(0.3~{\rm ms)}$, $\Delta_{33}/2\pi = 0.4 ~\rm MHz,$ 
$\Delta_{34}/\Delta_{33}=0.85$, $\Delta=\Delta_{33}/2$, $\Omega_R=2\Omega^2/\Delta_{33}$, and $\Omega_g=0$.  The coupling strengths between opposite corners were $\Delta_{33}'=\Delta_{33}/8$ and  $\Delta_{34}'=\Delta_{34}/8$. }
\label{fig.square}
\end{minipage}
\end{figure}

We demonstrate the ability of the Rydberg coupled system to find the ground state of the Ising model using numerical simulations.
The transverse field $B$ corresponds to $\Omega_g$ in (\ref{eq:Hamiltonian}). 
To find the quantity corresponding to $J$ we return to Eqs. (\ref{eq.rinout}).
In the limit of $\Delta_{33}\gg \gamma\gg \Omega_R$ the pumping rate into the AF state is dominated by 
$r_{2,\rm in}\simeq (1-P_{\rm AF})\frac{\Omega_R^2}{4\gamma}.$
We identify $r_{2,\rm in}\sim \frac{\Omega_R^2}{\gamma}$ with the spin-spin interaction strength $J$ in the Ising Hamiltonian. The numerical prefactor
is unknown as $P_{\rm AF}$ varies continuously during the dynamical evolution. 
Using an average value of $P_{\rm AF}=1/2$ we find $J= r_{2,\rm in}=
\frac{\Omega_R^2}{8\gamma}.$  The ratio of transverse field to spin-spin interaction strength $B/J$, which governs the nature of the ground state, thus maps onto the quantity $8\gamma\Omega_g/\Omega_R^2$. The assignment of the numerical prefactor  is verified by the good agreement with numerical  simulations in Fig. \ref{fig.square}.

Solving the master 
equation for more than several atoms requires large computational resources. We have instead used a Monte Carlo wavefunction technique\cite{Dalibard1992} to study the dynamics on three and four spin plaquettes. 
We will refer to internal  states $|1\rangle$, $|2\rangle$ as spin up and down. In  the case of three spins on an equilateral triangle the dynamics is frustrated and we find equal superpositions of the six distinct states with $M=\pm 1/2$ in the absence of a  transverse field.

Results for four atoms on a square plaquette are given in Fig. \ref{fig.square}. 
When there is no transverse field ($\Omega_g=0$) we pump close to 100\% of the population into a superposition of the $M=0$ states $|a\rangle=|1212\rangle$ and $|b\rangle=|2121\rangle$, corresponding to the ground state of the  Ising model (\ref{eq.Hising}) with $B=0$. The remaining few percent of the population is fairly evenly distributed over the 12 higher energy states. Solution of (\ref{eq.Hising}) on a square plaquette with $1/R^6$ van der Waals scaling of the interaction strength shows that the ground states have energy $\frac{-15J}{4}$, while the next four states are at $\frac{-J}{4}$. Our physical model with  Rydberg interactions and dissipation   shows relaxation to the ground state with purity of about 96\%. 

The population in the AF states $|a\rangle, |b\rangle$ in the presence of a transverse field is shown in the inset to the figure. The solid curve from numerical diagonalization of  (\ref{eq.Hising}) is shown with  
 points from Monte Carlo simulations of (\ref{eq.master}).
At large negative $B/J$ the ground state is $x$ polarized so the population in 
$\ket{a}$ and $\ket{b}$ asymptotes to 1/8. The good agreement between the curve and points demonstrates
the ability of the atomic model to relax to the ground state of the transverse field Ising model. 


In summary we have shown how the combination of weak Rydberg interactions and global rotations pump pairs of atoms into highly entangled singlet states. A related approach to entanglement generation using electromagnetically induced transparency with dissipative dynamics has been presented in 
\cite{Rao2013}. The entangled singlet  states can then be used as  a resource for high fidelity, long distance  gates in a qubit array based on gate teleportation\cite{Gottesman1999}. 

Going beyond pairs of atoms we show that the Ising model with transverse field can be mapped onto dissipative  Rydberg mediated dynamics. In the simplest nontrivial instance of long range interactions on a square plaquette we obtain good agreement 
for the antiferromagnetic fraction of the ground state found from the Ising Hamiltonian and from the atomic dynamics. The viability of the atomic interactions for simulating the ground states and dynamical evolution of Ising models on larger lattices and with inhomogeneous transverse fields  are open questions for future studies.

This work was supported by the IARPA MQCO program through  ARO contract W911NF-10-1-0347 and
by the NSF. MS thanks Klaus M\o{}lmer for a helpful discussion.

\bibliographystyle{apsrev4-1}


\begin{thebibliography}{48}%
\makeatletter
\providecommand \@ifxundefined [1]{%
 \@ifx{#1\undefined}
}%
\providecommand \@ifnum [1]{%
 \ifnum #1\expandafter \@firstoftwo
 \else \expandafter \@secondoftwo
 \fi
}%
\providecommand \@ifx [1]{%
 \ifx #1\expandafter \@firstoftwo
 \else \expandafter \@secondoftwo
 \fi
}%
\providecommand \natexlab [1]{#1}%
\providecommand \enquote  [1]{``#1''}%
\providecommand \bibnamefont  [1]{#1}%
\providecommand \bibfnamefont [1]{#1}%
\providecommand \citenamefont [1]{#1}%
\providecommand \href@noop [0]{\@secondoftwo}%
\providecommand \href [0]{\begingroup \@sanitize@url \@href}%
\providecommand \@href[1]{\@@startlink{#1}\@@href}%
\providecommand \@@href[1]{\endgroup#1\@@endlink}%
\providecommand \@sanitize@url [0]{\catcode `\\12\catcode `\$12\catcode
  `\&12\catcode `\#12\catcode `\^12\catcode `\_12\catcode `\%12\relax}%
\providecommand \@@startlink[1]{}%
\providecommand \@@endlink[0]{}%
\providecommand \url  [0]{\begingroup\@sanitize@url \@url }%
\providecommand \@url [1]{\endgroup\@href {#1}{\urlprefix }}%
\providecommand \urlprefix  [0]{URL }%
\providecommand \Eprint [0]{\href }%
\providecommand \doibase [0]{http://dx.doi.org/}%
\providecommand \selectlanguage [0]{\@gobble}%
\providecommand \bibinfo  [0]{\@secondoftwo}%
\providecommand \bibfield  [0]{\@secondoftwo}%
\providecommand \translation [1]{[#1]}%
\providecommand \BibitemOpen [0]{}%
\providecommand \bibitemStop [0]{}%
\providecommand \bibitemNoStop [0]{.\EOS\space}%
\providecommand \EOS [0]{\spacefactor3000\relax}%
\providecommand \BibitemShut  [1]{\csname bibitem#1\endcsname}%
\let\auto@bib@innerbib\@empty
\bibitem [{\citenamefont {Bloch}\ \emph {et~al.}(2008)\citenamefont {Bloch},
  \citenamefont {Dalibard},\ and\ \citenamefont {Zwerger}}]{Bloch2008b}%
  \BibitemOpen
  \bibfield  {author} {\bibinfo {author} {\bibfnamefont {I.}~\bibnamefont
  {Bloch}}, \bibinfo {author} {\bibfnamefont {J.}~\bibnamefont {Dalibard}}, \
  and\ \bibinfo {author} {\bibfnamefont {W.}~\bibnamefont {Zwerger}},\
  }\href@noop {} {\bibfield  {journal} {\bibinfo  {journal} {Rev. Mod. Phys.}\
  }\textbf {\bibinfo {volume} {80}},\ \bibinfo {pages} {885} (\bibinfo {year}
  {2008})}\BibitemShut {NoStop}%
\bibitem [{\citenamefont {Esslinger}(2010)}]{Esslinger2010}%
  \BibitemOpen
  \bibfield  {author} {\bibinfo {author} {\bibfnamefont {T.}~\bibnamefont
  {Esslinger}},\ }\href@noop {} {\bibfield  {journal} {\bibinfo  {journal}
  {Annu. Rev. Condens. Matter Phys.}\ }\textbf {\bibinfo {volume} {1}},\
  \bibinfo {pages} {129} (\bibinfo {year} {2010})}\BibitemShut {NoStop}%
\bibitem [{\citenamefont {Trefzger}\ \emph {et~al.}(2011)\citenamefont
  {Trefzger}, \citenamefont {Menotti}, \citenamefont {Capogrosso-Sansone},\
  and\ \citenamefont {Lewenstein}}]{Trefzger2011}%
  \BibitemOpen
  \bibfield  {author} {\bibinfo {author} {\bibfnamefont {C.}~\bibnamefont
  {Trefzger}}, \bibinfo {author} {\bibfnamefont {C.}~\bibnamefont {Menotti}},
  \bibinfo {author} {\bibfnamefont {B.}~\bibnamefont {Capogrosso-Sansone}}, \
  and\ \bibinfo {author} {\bibfnamefont {M.}~\bibnamefont {Lewenstein}},\
  }\href@noop {} {\bibfield  {journal} {\bibinfo  {journal} {J. Phys. B: At.
  Mol. Opt. Phys.}\ }\textbf {\bibinfo {volume} {44}},\ \bibinfo {pages}
  {193001} (\bibinfo {year} {2011})}\BibitemShut {NoStop}%
\bibitem [{\citenamefont {Baranov}\ \emph {et~al.}(2012)\citenamefont
  {Baranov}, \citenamefont {Dalmonte}, \citenamefont {Pupillo},\ and\
  \citenamefont {Zoller}}]{Baranov2012}%
  \BibitemOpen
  \bibfield  {author} {\bibinfo {author} {\bibfnamefont {M.~A.}\ \bibnamefont
  {Baranov}}, \bibinfo {author} {\bibfnamefont {M.}~\bibnamefont {Dalmonte}},
  \bibinfo {author} {\bibfnamefont {G.}~\bibnamefont {Pupillo}}, \ and\
  \bibinfo {author} {\bibfnamefont {P.}~\bibnamefont {Zoller}},\ }\href@noop {}
  {\bibfield  {journal} {\bibinfo  {journal} {Chem. Rev.}\ }\textbf {\bibinfo
  {volume} {112}},\ \bibinfo {pages} {5012} (\bibinfo {year}
  {2012})}\BibitemShut {NoStop}%
\bibitem [{\citenamefont {Simon}\ \emph {et~al.}(2011)\citenamefont {Simon},
  \citenamefont {Bakr}, \citenamefont {Ma}, \citenamefont {Tai}, \citenamefont
  {Preiss},\ and\ \citenamefont {Greiner}}]{Simon2011}%
  \BibitemOpen
  \bibfield  {author} {\bibinfo {author} {\bibfnamefont {J.}~\bibnamefont
  {Simon}}, \bibinfo {author} {\bibfnamefont {W.~S.}\ \bibnamefont {Bakr}},
  \bibinfo {author} {\bibfnamefont {R.}~\bibnamefont {Ma}}, \bibinfo {author}
  {\bibfnamefont {M.~E.}\ \bibnamefont {Tai}}, \bibinfo {author} {\bibfnamefont
  {P.~M.}\ \bibnamefont {Preiss}}, \ and\ \bibinfo {author} {\bibfnamefont
  {M.}~\bibnamefont {Greiner}},\ }\href@noop {} {\bibfield  {journal} {\bibinfo
   {journal} {Nature}\ }\textbf {\bibinfo {volume} {472}},\ \bibinfo {pages}
  {307} (\bibinfo {year} {2011})}\BibitemShut {NoStop}%
\bibitem [{\citenamefont {Zhang}\ \emph
  {et~al.}(2012{\natexlab{a}})\citenamefont {Zhang}, \citenamefont {Hung},
  \citenamefont {Tung},\ and\ \citenamefont {Chin}}]{Zhang2012}%
  \BibitemOpen
  \bibfield  {author} {\bibinfo {author} {\bibfnamefont {X.}~\bibnamefont
  {Zhang}}, \bibinfo {author} {\bibfnamefont {C.-L.}\ \bibnamefont {Hung}},
  \bibinfo {author} {\bibfnamefont {S.-K.}\ \bibnamefont {Tung}}, \ and\
  \bibinfo {author} {\bibfnamefont {C.}~\bibnamefont {Chin}},\ }\href@noop {}
  {\bibfield  {journal} {\bibinfo  {journal} {Science}\ }\textbf {\bibinfo
  {volume} {335}},\ \bibinfo {pages} {1070} (\bibinfo {year}
  {2012}{\natexlab{a}})}\BibitemShut {NoStop}%
\bibitem [{\citenamefont {Friedenauer}\ \emph {et~al.}(2008)\citenamefont
  {Friedenauer}, \citenamefont {Schmitz}, \citenamefont {Glueckert},
  \citenamefont {Porras},\ and\ \citenamefont {Schaetz}}]{Friedenauer2010}%
  \BibitemOpen
  \bibfield  {author} {\bibinfo {author} {\bibfnamefont {A.}~\bibnamefont
  {Friedenauer}}, \bibinfo {author} {\bibfnamefont {H.}~\bibnamefont
  {Schmitz}}, \bibinfo {author} {\bibfnamefont {J.~T.}\ \bibnamefont
  {Glueckert}}, \bibinfo {author} {\bibfnamefont {D.}~\bibnamefont {Porras}}, \
  and\ \bibinfo {author} {\bibfnamefont {T.}~\bibnamefont {Schaetz}},\
  }\href@noop {} {\bibfield  {journal} {\bibinfo  {journal} {Nat. Phys.}\
  }\textbf {\bibinfo {volume} {4}},\ \bibinfo {pages} {757} (\bibinfo {year}
  {2008})}\BibitemShut {NoStop}%
\bibitem [{\citenamefont {Kim}\ \emph {et~al.}(2010)\citenamefont {Kim},
  \citenamefont {Chang}, \citenamefont {Korenblit}, \citenamefont {Islam},
  \citenamefont {Edwards}, \citenamefont {Freericks}, \citenamefont {Lin},
  \citenamefont {Duan},\ and\ \citenamefont {Monroe}}]{Kim2010}%
  \BibitemOpen
  \bibfield  {author} {\bibinfo {author} {\bibfnamefont {K.}~\bibnamefont
  {Kim}}, \bibinfo {author} {\bibfnamefont {M.-S.}\ \bibnamefont {Chang}},
  \bibinfo {author} {\bibfnamefont {S.}~\bibnamefont {Korenblit}}, \bibinfo
  {author} {\bibfnamefont {R.}~\bibnamefont {Islam}}, \bibinfo {author}
  {\bibfnamefont {E.~E.}\ \bibnamefont {Edwards}}, \bibinfo {author}
  {\bibfnamefont {J.~K.}\ \bibnamefont {Freericks}}, \bibinfo {author}
  {\bibfnamefont {G.-D.}\ \bibnamefont {Lin}}, \bibinfo {author} {\bibfnamefont
  {L.-M.}\ \bibnamefont {Duan}}, \ and\ \bibinfo {author} {\bibfnamefont
  {C.}~\bibnamefont {Monroe}},\ }\href@noop {} {\bibfield  {journal} {\bibinfo
  {journal} {Nature}\ }\textbf {\bibinfo {volume} {465}},\ \bibinfo {pages}
  {590} (\bibinfo {year} {2010})}\BibitemShut {NoStop}%
\bibitem [{\citenamefont {Lanyon}\ \emph {et~al.}(2011)\citenamefont {Lanyon},
  \citenamefont {Hempel}, \citenamefont {Nigg}, \citenamefont {M\"uller},
  \citenamefont {Gerritsma}, \citenamefont {Z\"ahringer}, \citenamefont
  {Schindler}, \citenamefont {Barreiro}, \citenamefont {Rambach}, \citenamefont
  {Kirchmair}, \citenamefont {Hennrich}, \citenamefont {Zoller}, \citenamefont
  {Blatt},\ and\ \citenamefont {Roos}}]{Lanyon2011}%
  \BibitemOpen
  \bibfield  {author} {\bibinfo {author} {\bibfnamefont {B.~P.}\ \bibnamefont
  {Lanyon}}, \bibinfo {author} {\bibfnamefont {C.}~\bibnamefont {Hempel}},
  \bibinfo {author} {\bibfnamefont {D.}~\bibnamefont {Nigg}}, \bibinfo {author}
  {\bibfnamefont {M.}~\bibnamefont {M\"uller}}, \bibinfo {author}
  {\bibfnamefont {R.}~\bibnamefont {Gerritsma}}, \bibinfo {author}
  {\bibfnamefont {F.}~\bibnamefont {Z\"ahringer}}, \bibinfo {author}
  {\bibfnamefont {P.}~\bibnamefont {Schindler}}, \bibinfo {author}
  {\bibfnamefont {J.~T.}\ \bibnamefont {Barreiro}}, \bibinfo {author}
  {\bibfnamefont {M.}~\bibnamefont {Rambach}}, \bibinfo {author} {\bibfnamefont
  {G.}~\bibnamefont {Kirchmair}}, \bibinfo {author} {\bibfnamefont
  {M.}~\bibnamefont {Hennrich}}, \bibinfo {author} {\bibfnamefont
  {P.}~\bibnamefont {Zoller}}, \bibinfo {author} {\bibfnamefont
  {R.}~\bibnamefont {Blatt}}, \ and\ \bibinfo {author} {\bibfnamefont {C.~F.}\
  \bibnamefont {Roos}},\ }\href@noop {} {\bibfield  {journal} {\bibinfo
  {journal} {Science}\ }\textbf {\bibinfo {volume} {334}},\ \bibinfo {pages}
  {57} (\bibinfo {year} {2011})}\BibitemShut {NoStop}%
\bibitem [{\citenamefont {Britton}\ \emph {et~al.}(2012)\citenamefont
  {Britton}, \citenamefont {Sawyer}, \citenamefont {Keith}, \citenamefont
  {Wang}, \citenamefont {Freericks}, \citenamefont {Uys}, \citenamefont
  {Biercuk},\ and\ \citenamefont {Bollinger}}]{Britton2012}%
  \BibitemOpen
  \bibfield  {author} {\bibinfo {author} {\bibfnamefont {J.~W.}\ \bibnamefont
  {Britton}}, \bibinfo {author} {\bibfnamefont {B.~C.}\ \bibnamefont {Sawyer}},
  \bibinfo {author} {\bibfnamefont {A.~C.}\ \bibnamefont {Keith}}, \bibinfo
  {author} {\bibfnamefont {C.-C.~J.}\ \bibnamefont {Wang}}, \bibinfo {author}
  {\bibfnamefont {J.~K.}\ \bibnamefont {Freericks}}, \bibinfo {author}
  {\bibfnamefont {H.}~\bibnamefont {Uys}}, \bibinfo {author} {\bibfnamefont
  {M.~J.}\ \bibnamefont {Biercuk}}, \ and\ \bibinfo {author} {\bibfnamefont
  {J.~J.}\ \bibnamefont {Bollinger}},\ }\href@noop {} {\bibfield  {journal}
  {\bibinfo  {journal} {Nature}\ }\textbf {\bibinfo {volume} {484}},\ \bibinfo
  {pages} {489} (\bibinfo {year} {2012})}\BibitemShut {NoStop}%
\bibitem [{\citenamefont {J\"ordens}\ \emph {et~al.}(2010)\citenamefont
  {J\"ordens}, \citenamefont {Tarruell}, \citenamefont {Greif}, \citenamefont
  {Uehlinger}, \citenamefont {Strohmaier}, \citenamefont {Moritz},
  \citenamefont {Esslinger}, \citenamefont {De~Leo}, \citenamefont {Kollath},
  \citenamefont {Georges}, \citenamefont {Scarola}, \citenamefont {Pollet},
  \citenamefont {Burovski}, \citenamefont {Kozik},\ and\ \citenamefont
  {Troyer}}]{Jordens2010}%
  \BibitemOpen
  \bibfield  {author} {\bibinfo {author} {\bibfnamefont {R.}~\bibnamefont
  {J\"ordens}}, \bibinfo {author} {\bibfnamefont {L.}~\bibnamefont {Tarruell}},
  \bibinfo {author} {\bibfnamefont {D.}~\bibnamefont {Greif}}, \bibinfo
  {author} {\bibfnamefont {T.}~\bibnamefont {Uehlinger}}, \bibinfo {author}
  {\bibfnamefont {N.}~\bibnamefont {Strohmaier}}, \bibinfo {author}
  {\bibfnamefont {H.}~\bibnamefont {Moritz}}, \bibinfo {author} {\bibfnamefont
  {T.}~\bibnamefont {Esslinger}}, \bibinfo {author} {\bibfnamefont
  {L.}~\bibnamefont {De~Leo}}, \bibinfo {author} {\bibfnamefont
  {C.}~\bibnamefont {Kollath}}, \bibinfo {author} {\bibfnamefont
  {A.}~\bibnamefont {Georges}}, \bibinfo {author} {\bibfnamefont
  {V.}~\bibnamefont {Scarola}}, \bibinfo {author} {\bibfnamefont
  {L.}~\bibnamefont {Pollet}}, \bibinfo {author} {\bibfnamefont
  {E.}~\bibnamefont {Burovski}}, \bibinfo {author} {\bibfnamefont
  {E.}~\bibnamefont {Kozik}}, \ and\ \bibinfo {author} {\bibfnamefont
  {M.}~\bibnamefont {Troyer}},\ }\href@noop {} {\bibfield  {journal} {\bibinfo
  {journal} {Phys. Rev. Lett.}\ }\textbf {\bibinfo {volume} {104}},\ \bibinfo
  {pages} {180401} (\bibinfo {year} {2010})}\BibitemShut {NoStop}%
\bibitem [{\citenamefont {Saffman}\ \emph {et~al.}(2010)\citenamefont
  {Saffman}, \citenamefont {Walker},\ and\ \citenamefont
  {M\o{}lmer}}]{Saffman2010}%
  \BibitemOpen
  \bibfield  {author} {\bibinfo {author} {\bibfnamefont {M.}~\bibnamefont
  {Saffman}}, \bibinfo {author} {\bibfnamefont {T.~G.}\ \bibnamefont {Walker}},
  \ and\ \bibinfo {author} {\bibfnamefont {K.}~\bibnamefont {M\o{}lmer}},\
  }\href@noop {} {\bibfield  {journal} {\bibinfo  {journal} {Rev. Mod. Phys.}\
  }\textbf {\bibinfo {volume} {82}},\ \bibinfo {pages} {2313} (\bibinfo {year}
  {2010})}\BibitemShut {NoStop}%
\bibitem [{\citenamefont {L\"ow}\ \emph {et~al.}(2012)\citenamefont {L\"ow},
  \citenamefont {Weimer}, \citenamefont {Nipper}, \citenamefont {Balewski},
  \citenamefont {Butscher}, \citenamefont {B\"uchler},\ and\ \citenamefont
  {Pfau}}]{Low2012}%
  \BibitemOpen
  \bibfield  {author} {\bibinfo {author} {\bibfnamefont {R.}~\bibnamefont
  {L\"ow}}, \bibinfo {author} {\bibfnamefont {H.}~\bibnamefont {Weimer}},
  \bibinfo {author} {\bibfnamefont {J.}~\bibnamefont {Nipper}}, \bibinfo
  {author} {\bibfnamefont {J.~B.}\ \bibnamefont {Balewski}}, \bibinfo {author}
  {\bibfnamefont {B.}~\bibnamefont {Butscher}}, \bibinfo {author}
  {\bibfnamefont {H.~P.}\ \bibnamefont {B\"uchler}}, \ and\ \bibinfo {author}
  {\bibfnamefont {T.}~\bibnamefont {Pfau}},\ }\href@noop {} {\bibfield
  {journal} {\bibinfo  {journal} {J. Phys. B: At. Mol. Opt. Phys.}\ }\textbf
  {\bibinfo {volume} {45}},\ \bibinfo {pages} {113001} (\bibinfo {year}
  {2012})}\BibitemShut {NoStop}%
\bibitem [{\citenamefont {Weimer}\ and\ \citenamefont
  {B\"uchler}(2010)}]{Weimer2010b}%
  \BibitemOpen
  \bibfield  {author} {\bibinfo {author} {\bibfnamefont {H.}~\bibnamefont
  {Weimer}}\ and\ \bibinfo {author} {\bibfnamefont {H.~P.}\ \bibnamefont
  {B\"uchler}},\ }\href@noop {} {\bibfield  {journal} {\bibinfo  {journal}
  {Phys. Rev. Lett.}\ }\textbf {\bibinfo {volume} {105}},\ \bibinfo {pages}
  {230403} (\bibinfo {year} {2010})}\BibitemShut {NoStop}%
\bibitem [{\citenamefont {Weimer}\ \emph {et~al.}(2010)\citenamefont {Weimer},
  \citenamefont {M\"uller}, \citenamefont {Lesanovsky}, \citenamefont
  {Zoller},\ and\ \citenamefont {B\"uchler}}]{Weimer2010}%
  \BibitemOpen
  \bibfield  {author} {\bibinfo {author} {\bibfnamefont {H.}~\bibnamefont
  {Weimer}}, \bibinfo {author} {\bibfnamefont {M.}~\bibnamefont {M\"uller}},
  \bibinfo {author} {\bibfnamefont {I.}~\bibnamefont {Lesanovsky}}, \bibinfo
  {author} {\bibfnamefont {P.}~\bibnamefont {Zoller}}, \ and\ \bibinfo {author}
  {\bibfnamefont {H.~P.}\ \bibnamefont {B\"uchler}},\ }\href@noop {} {\bibfield
   {journal} {\bibinfo  {journal} {Nat. Phys.}\ }\textbf {\bibinfo {volume}
  {6}},\ \bibinfo {pages} {382} (\bibinfo {year} {2010})}\BibitemShut {NoStop}%
\bibitem [{\citenamefont {Sela}\ \emph {et~al.}(2011)\citenamefont {Sela},
  \citenamefont {Punk},\ and\ \citenamefont {Garst}}]{Sela2011}%
  \BibitemOpen
  \bibfield  {author} {\bibinfo {author} {\bibfnamefont {E.}~\bibnamefont
  {Sela}}, \bibinfo {author} {\bibfnamefont {M.}~\bibnamefont {Punk}}, \ and\
  \bibinfo {author} {\bibfnamefont {M.}~\bibnamefont {Garst}},\ }\href@noop {}
  {\bibfield  {journal} {\bibinfo  {journal} {Phys. Rev. B}\ }\textbf {\bibinfo
  {volume} {84}},\ \bibinfo {pages} {085434} (\bibinfo {year}
  {2011})}\BibitemShut {NoStop}%
\bibitem [{\citenamefont {Lesanovsky}(2011)}]{Lesanovsky2011}%
  \BibitemOpen
  \bibfield  {author} {\bibinfo {author} {\bibfnamefont {I.}~\bibnamefont
  {Lesanovsky}},\ }\href@noop {} {\bibfield  {journal} {\bibinfo  {journal}
  {Phys. Rev. Lett.}\ }\textbf {\bibinfo {volume} {106}},\ \bibinfo {pages}
  {025301} (\bibinfo {year} {2011})}\BibitemShut {NoStop}%
\bibitem [{\citenamefont {Ji}\ \emph {et~al.}(2011)\citenamefont {Ji},
  \citenamefont {Ates},\ and\ \citenamefont {Lesanovsky}}]{Ji2011}%
  \BibitemOpen
  \bibfield  {author} {\bibinfo {author} {\bibfnamefont {S.}~\bibnamefont
  {Ji}}, \bibinfo {author} {\bibfnamefont {C.}~\bibnamefont {Ates}}, \ and\
  \bibinfo {author} {\bibfnamefont {I.}~\bibnamefont {Lesanovsky}},\
  }\href@noop {} {\bibfield  {journal} {\bibinfo  {journal} {Phys. Rev. Lett.}\
  }\textbf {\bibinfo {volume} {107}},\ \bibinfo {pages} {060406} (\bibinfo
  {year} {2011})}\BibitemShut {NoStop}%
\bibitem [{\citenamefont {Zeller}\ \emph {et~al.}(2012)\citenamefont {Zeller},
  \citenamefont {Mayle}, \citenamefont {Bonato}, \citenamefont {Reinelt},\ and\
  \citenamefont {Schmelcher}}]{Zeller2012}%
  \BibitemOpen
  \bibfield  {author} {\bibinfo {author} {\bibfnamefont {W.}~\bibnamefont
  {Zeller}}, \bibinfo {author} {\bibfnamefont {M.}~\bibnamefont {Mayle}},
  \bibinfo {author} {\bibfnamefont {T.}~\bibnamefont {Bonato}}, \bibinfo
  {author} {\bibfnamefont {G.}~\bibnamefont {Reinelt}}, \ and\ \bibinfo
  {author} {\bibfnamefont {P.}~\bibnamefont {Schmelcher}},\ }\href@noop {}
  {\bibfield  {journal} {\bibinfo  {journal} {Phys. Rev. A}\ }\textbf {\bibinfo
  {volume} {85}},\ \bibinfo {pages} {063603} (\bibinfo {year}
  {2012})}\BibitemShut {NoStop}%
\bibitem [{\citenamefont {Plenio}\ \emph {et~al.}(1999)\citenamefont {Plenio},
  \citenamefont {Huelga}, \citenamefont {Beige},\ and\ \citenamefont
  {Knight}}]{Plenio1999}%
  \BibitemOpen
  \bibfield  {author} {\bibinfo {author} {\bibfnamefont {M.~B.}\ \bibnamefont
  {Plenio}}, \bibinfo {author} {\bibfnamefont {S.~F.}\ \bibnamefont {Huelga}},
  \bibinfo {author} {\bibfnamefont {A.}~\bibnamefont {Beige}}, \ and\ \bibinfo
  {author} {\bibfnamefont {P.~L.}\ \bibnamefont {Knight}},\ }\href@noop {}
  {\bibfield  {journal} {\bibinfo  {journal} {Phys. Rev. A}\ }\textbf {\bibinfo
  {volume} {59}},\ \bibinfo {pages} {2468} (\bibinfo {year}
  {1999})}\BibitemShut {NoStop}%
\bibitem [{\citenamefont {Schneider}\ and\ \citenamefont
  {Milburn}(2002)}]{Schneider2002}%
  \BibitemOpen
  \bibfield  {author} {\bibinfo {author} {\bibfnamefont {S.}~\bibnamefont
  {Schneider}}\ and\ \bibinfo {author} {\bibfnamefont {G.~J.}\ \bibnamefont
  {Milburn}},\ }\href@noop {} {\bibfield  {journal} {\bibinfo  {journal} {Phys.
  Rev. A}\ }\textbf {\bibinfo {volume} {65}},\ \bibinfo {pages} {042107}
  (\bibinfo {year} {2002})}\BibitemShut {NoStop}%
\bibitem [{\citenamefont {Braun}(2002)}]{Braun2002}%
  \BibitemOpen
  \bibfield  {author} {\bibinfo {author} {\bibfnamefont {D.}~\bibnamefont
  {Braun}},\ }\href@noop {} {\bibfield  {journal} {\bibinfo  {journal} {Phys.
  Rev. Lett.}\ }\textbf {\bibinfo {volume} {89}},\ \bibinfo {pages} {277901}
  (\bibinfo {year} {2002})}\BibitemShut {NoStop}%
\bibitem [{\citenamefont {Jak\'obczyk}(2002)}]{Jakobczyk2002}%
  \BibitemOpen
  \bibfield  {author} {\bibinfo {author} {\bibfnamefont {L.}~\bibnamefont
  {Jak\'obczyk}},\ }\href@noop {} {\bibfield  {journal} {\bibinfo  {journal}
  {J. Phys. A}\ }\textbf {\bibinfo {volume} {35}},\ \bibinfo {pages} {6383}
  (\bibinfo {year} {2002})}\BibitemShut {NoStop}%
\bibitem [{\citenamefont {Basharov}(2002)}]{Basharov2002}%
  \BibitemOpen
  \bibfield  {author} {\bibinfo {author} {\bibfnamefont {A.~M.}\ \bibnamefont
  {Basharov}},\ }\href@noop {} {\bibfield  {journal} {\bibinfo  {journal}
  {Pisma Zh. \'Eksp. Teor. Fiz.}\ }\textbf {\bibinfo {volume} {75}},\ \bibinfo
  {pages} {151} (\bibinfo {year} {2002})},\ \bibinfo {note} {[JETP Lett. {\bf
  75}, 123 (2002)]}\BibitemShut {NoStop}%
\bibitem [{\citenamefont {Diehl}\ \emph {et~al.}(2008)\citenamefont {Diehl},
  \citenamefont {Micheli}, \citenamefont {Kantian}, \citenamefont {Kraus},
  \citenamefont {B\"uchler},\ and\ \citenamefont {Zoller}}]{Diehl2008}%
  \BibitemOpen
  \bibfield  {author} {\bibinfo {author} {\bibfnamefont {S.}~\bibnamefont
  {Diehl}}, \bibinfo {author} {\bibfnamefont {A.}~\bibnamefont {Micheli}},
  \bibinfo {author} {\bibfnamefont {A.}~\bibnamefont {Kantian}}, \bibinfo
  {author} {\bibfnamefont {B.}~\bibnamefont {Kraus}}, \bibinfo {author}
  {\bibfnamefont {H.~P.}\ \bibnamefont {B\"uchler}}, \ and\ \bibinfo {author}
  {\bibfnamefont {P.}~\bibnamefont {Zoller}},\ }\href@noop {} {\bibfield
  {journal} {\bibinfo  {journal} {Nat. Phys.}\ }\textbf {\bibinfo {volume}
  {4}},\ \bibinfo {pages} {878} (\bibinfo {year} {2008})}\BibitemShut {NoStop}%
\bibitem [{\citenamefont {Verstraete}\ \emph {et~al.}(2009)\citenamefont
  {Verstraete}, \citenamefont {Wolf},\ and\ \citenamefont
  {Cirac}}]{Verstraete2009}%
  \BibitemOpen
  \bibfield  {author} {\bibinfo {author} {\bibfnamefont {F.}~\bibnamefont
  {Verstraete}}, \bibinfo {author} {\bibfnamefont {M.~M.}\ \bibnamefont
  {Wolf}}, \ and\ \bibinfo {author} {\bibfnamefont {J.~I.}\ \bibnamefont
  {Cirac}},\ }\href@noop {} {\bibfield  {journal} {\bibinfo  {journal} {Nat.
  Phys.}\ }\textbf {\bibinfo {volume} {5}},\ \bibinfo {pages} {633} (\bibinfo
  {year} {2009})}\BibitemShut {NoStop}%
\bibitem [{\citenamefont {Krauter}\ \emph {et~al.}(2011)\citenamefont
  {Krauter}, \citenamefont {Muschik}, \citenamefont {Jensen}, \citenamefont
  {Wasilewski}, \citenamefont {Petersen}, \citenamefont {Cirac},\ and\
  \citenamefont {Polzik}}]{Krauter2011}%
  \BibitemOpen
  \bibfield  {author} {\bibinfo {author} {\bibfnamefont {H.}~\bibnamefont
  {Krauter}}, \bibinfo {author} {\bibfnamefont {C.~A.}\ \bibnamefont
  {Muschik}}, \bibinfo {author} {\bibfnamefont {K.}~\bibnamefont {Jensen}},
  \bibinfo {author} {\bibfnamefont {W.}~\bibnamefont {Wasilewski}}, \bibinfo
  {author} {\bibfnamefont {J.~M.}\ \bibnamefont {Petersen}}, \bibinfo {author}
  {\bibfnamefont {J.~I.}\ \bibnamefont {Cirac}}, \ and\ \bibinfo {author}
  {\bibfnamefont {E.~S.}\ \bibnamefont {Polzik}},\ }\href@noop {} {\bibfield
  {journal} {\bibinfo  {journal} {Phys. Rev. Lett.}\ }\textbf {\bibinfo
  {volume} {107}},\ \bibinfo {pages} {080503} (\bibinfo {year}
  {2011})}\BibitemShut {NoStop}%
\bibitem [{\citenamefont {Barreiro}\ \emph {et~al.}(2011)\citenamefont
  {Barreiro}, \citenamefont {M\"uller}, \citenamefont {Schindler},
  \citenamefont {Nigg}, \citenamefont {Monz}, \citenamefont {Chwalla},
  \citenamefont {Hennrich}, \citenamefont {Roos}, \citenamefont {Zoller},\ and\
  \citenamefont {Blatt}}]{Barreiro2011}%
  \BibitemOpen
  \bibfield  {author} {\bibinfo {author} {\bibfnamefont {J.~T.}\ \bibnamefont
  {Barreiro}}, \bibinfo {author} {\bibfnamefont {M.}~\bibnamefont {M\"uller}},
  \bibinfo {author} {\bibfnamefont {P.}~\bibnamefont {Schindler}}, \bibinfo
  {author} {\bibfnamefont {D.}~\bibnamefont {Nigg}}, \bibinfo {author}
  {\bibfnamefont {T.}~\bibnamefont {Monz}}, \bibinfo {author} {\bibfnamefont
  {M.}~\bibnamefont {Chwalla}}, \bibinfo {author} {\bibfnamefont
  {M.}~\bibnamefont {Hennrich}}, \bibinfo {author} {\bibfnamefont {C.~F.}\
  \bibnamefont {Roos}}, \bibinfo {author} {\bibfnamefont {P.}~\bibnamefont
  {Zoller}}, \ and\ \bibinfo {author} {\bibfnamefont {R.}~\bibnamefont
  {Blatt}},\ }\href@noop {} {\bibfield  {journal} {\bibinfo  {journal}
  {Nature}\ }\textbf {\bibinfo {volume} {470}},\ \bibinfo {pages} {486}
  (\bibinfo {year} {2011})}\BibitemShut {NoStop}%
\bibitem [{\citenamefont {Mart\'in-Cano}\ \emph {et~al.}(2011)\citenamefont
  {Mart\'in-Cano}, \citenamefont {Gonz\'alez-Tudela}, \citenamefont
  {Mart\'in-Moreno}, \citenamefont {Garc\'ia-Vidal}, \citenamefont {Tejedor},\
  and\ \citenamefont {Moreno}}]{Martin-Cano2011}%
  \BibitemOpen
  \bibfield  {author} {\bibinfo {author} {\bibfnamefont {D.}~\bibnamefont
  {Mart\'in-Cano}}, \bibinfo {author} {\bibfnamefont {A.}~\bibnamefont
  {Gonz\'alez-Tudela}}, \bibinfo {author} {\bibfnamefont {L.}~\bibnamefont
  {Mart\'in-Moreno}}, \bibinfo {author} {\bibfnamefont {F.~J.}\ \bibnamefont
  {Garc\'ia-Vidal}}, \bibinfo {author} {\bibfnamefont {C.}~\bibnamefont
  {Tejedor}}, \ and\ \bibinfo {author} {\bibfnamefont {E.}~\bibnamefont
  {Moreno}},\ }\href@noop {} {\bibfield  {journal} {\bibinfo  {journal} {Phys.
  Rev. B}\ }\textbf {\bibinfo {volume} {84}},\ \bibinfo {pages} {235306}
  (\bibinfo {year} {2011})}\BibitemShut {NoStop}%
\bibitem [{\citenamefont {Murch}\ \emph {et~al.}(2012)\citenamefont {Murch},
  \citenamefont {Vool}, \citenamefont {Zhou}, \citenamefont {Weber},
  \citenamefont {Girvin},\ and\ \citenamefont {Siddiqi}}]{Murch2012}%
  \BibitemOpen
  \bibfield  {author} {\bibinfo {author} {\bibfnamefont {K.~W.}\ \bibnamefont
  {Murch}}, \bibinfo {author} {\bibfnamefont {U.}~\bibnamefont {Vool}},
  \bibinfo {author} {\bibfnamefont {D.}~\bibnamefont {Zhou}}, \bibinfo {author}
  {\bibfnamefont {S.~J.}\ \bibnamefont {Weber}}, \bibinfo {author}
  {\bibfnamefont {S.~M.}\ \bibnamefont {Girvin}}, \ and\ \bibinfo {author}
  {\bibfnamefont {I.}~\bibnamefont {Siddiqi}},\ }\href@noop {} {\bibfield
  {journal} {\bibinfo  {journal} {Phys. Rev. Lett.}\ }\textbf {\bibinfo
  {volume} {109}},\ \bibinfo {pages} {183602} (\bibinfo {year}
  {2012})}\BibitemShut {NoStop}%
\bibitem [{\citenamefont {Cho}\ \emph {et~al.}(2011)\citenamefont {Cho},
  \citenamefont {Bose},\ and\ \citenamefont {Kim}}]{Cho2011}%
  \BibitemOpen
  \bibfield  {author} {\bibinfo {author} {\bibfnamefont {J.}~\bibnamefont
  {Cho}}, \bibinfo {author} {\bibfnamefont {S.}~\bibnamefont {Bose}}, \ and\
  \bibinfo {author} {\bibfnamefont {M.~S.}\ \bibnamefont {Kim}},\ }\href@noop
  {} {\bibfield  {journal} {\bibinfo  {journal} {Phys. Rev. Lett.}\ }\textbf
  {\bibinfo {volume} {106}},\ \bibinfo {pages} {020504} (\bibinfo {year}
  {2011})}\BibitemShut {NoStop}%
\bibitem [{\citenamefont {Kordas}\ \emph {et~al.}(2012)\citenamefont {Kordas},
  \citenamefont {Wimberger},\ and\ \citenamefont {Witthaut}}]{Kordas2012}%
  \BibitemOpen
  \bibfield  {author} {\bibinfo {author} {\bibfnamefont {G.}~\bibnamefont
  {Kordas}}, \bibinfo {author} {\bibfnamefont {S.}~\bibnamefont {Wimberger}}, \
  and\ \bibinfo {author} {\bibfnamefont {D.}~\bibnamefont {Witthaut}},\
  }\href@noop {} {\bibfield  {journal} {\bibinfo  {journal} {Eur. Phys. Lett.}\
  }\textbf {\bibinfo {volume} {100}},\ \bibinfo {pages} {30007} (\bibinfo
  {year} {2012})}\BibitemShut {NoStop}%
\bibitem [{\citenamefont {Dalla~Torre}\ \emph {et~al.}(2013)\citenamefont
  {Dalla~Torre}, \citenamefont {Otterbach}, \citenamefont {Demler},
  \citenamefont {Vuletic},\ and\ \citenamefont {Lukin}}]{DallaTorre2013}%
  \BibitemOpen
  \bibfield  {author} {\bibinfo {author} {\bibfnamefont {E.~G.}\ \bibnamefont
  {Dalla~Torre}}, \bibinfo {author} {\bibfnamefont {J.}~\bibnamefont
  {Otterbach}}, \bibinfo {author} {\bibfnamefont {E.}~\bibnamefont {Demler}},
  \bibinfo {author} {\bibfnamefont {V.}~\bibnamefont {Vuletic}}, \ and\
  \bibinfo {author} {\bibfnamefont {M.~D.}\ \bibnamefont {Lukin}},\ }\href@noop
  {} {\bibfield  {journal} {\bibinfo  {journal} {Phys. Rev. Lett.}\ }\textbf
  {\bibinfo {volume} {110}},\ \bibinfo {pages} {120402} (\bibinfo {year}
  {2013})}\BibitemShut {NoStop}%
\bibitem [{\citenamefont {Tan}\ \emph {et~al.}(2013)\citenamefont {Tan},
  \citenamefont {Li},\ and\ \citenamefont {Meystre}}]{Tan2013}%
  \BibitemOpen
  \bibfield  {author} {\bibinfo {author} {\bibfnamefont {H.}~\bibnamefont
  {Tan}}, \bibinfo {author} {\bibfnamefont {G.}~\bibnamefont {Li}}, \ and\
  \bibinfo {author} {\bibfnamefont {P.}~\bibnamefont {Meystre}},\ }\href@noop
  {} {\bibfield  {journal} {\bibinfo  {journal} {Phys. Rev. A}\ }\textbf
  {\bibinfo {volume} {87}},\ \bibinfo {pages} {033829} (\bibinfo {year}
  {2013})}\BibitemShut {NoStop}%
\bibitem [{\citenamefont {Lee}\ \emph {et~al.}(2011)\citenamefont {Lee},
  \citenamefont {H\"affner},\ and\ \citenamefont {Cross}}]{Lee2011}%
  \BibitemOpen
  \bibfield  {author} {\bibinfo {author} {\bibfnamefont {T.~E.}\ \bibnamefont
  {Lee}}, \bibinfo {author} {\bibfnamefont {H.}~\bibnamefont {H\"affner}}, \
  and\ \bibinfo {author} {\bibfnamefont {M.~C.}\ \bibnamefont {Cross}},\
  }\href@noop {} {\bibfield  {journal} {\bibinfo  {journal} {Phys. Rev. A}\
  }\textbf {\bibinfo {volume} {84}},\ \bibinfo {pages} {031402(R)} (\bibinfo
  {year} {2011})}\BibitemShut {NoStop}%
\bibitem [{\citenamefont {Wilk}\ \emph {et~al.}(2010)\citenamefont {Wilk},
  \citenamefont {Ga\"etan}, \citenamefont {Evellin}, \citenamefont {Wolters},
  \citenamefont {Miroshnychenko}, \citenamefont {Grangier},\ and\ \citenamefont
  {Browaeys}}]{Wilk2010}%
  \BibitemOpen
  \bibfield  {author} {\bibinfo {author} {\bibfnamefont {T.}~\bibnamefont
  {Wilk}}, \bibinfo {author} {\bibfnamefont {A.}~\bibnamefont {Ga\"etan}},
  \bibinfo {author} {\bibfnamefont {C.}~\bibnamefont {Evellin}}, \bibinfo
  {author} {\bibfnamefont {J.}~\bibnamefont {Wolters}}, \bibinfo {author}
  {\bibfnamefont {Y.}~\bibnamefont {Miroshnychenko}}, \bibinfo {author}
  {\bibfnamefont {P.}~\bibnamefont {Grangier}}, \ and\ \bibinfo {author}
  {\bibfnamefont {A.}~\bibnamefont {Browaeys}},\ }\href@noop {} {\bibfield
  {journal} {\bibinfo  {journal} {Phys. Rev. Lett.}\ }\textbf {\bibinfo
  {volume} {104}},\ \bibinfo {pages} {010502} (\bibinfo {year}
  {2010})}\BibitemShut {NoStop}%
\bibitem [{\citenamefont {Isenhower}\ \emph {et~al.}(2010)\citenamefont
  {Isenhower}, \citenamefont {Urban}, \citenamefont {Zhang}, \citenamefont
  {Gill}, \citenamefont {Henage}, \citenamefont {Johnson}, \citenamefont
  {Walker},\ and\ \citenamefont {Saffman}}]{Isenhower2010}%
  \BibitemOpen
  \bibfield  {author} {\bibinfo {author} {\bibfnamefont {L.}~\bibnamefont
  {Isenhower}}, \bibinfo {author} {\bibfnamefont {E.}~\bibnamefont {Urban}},
  \bibinfo {author} {\bibfnamefont {X.~L.}\ \bibnamefont {Zhang}}, \bibinfo
  {author} {\bibfnamefont {A.~T.}\ \bibnamefont {Gill}}, \bibinfo {author}
  {\bibfnamefont {T.}~\bibnamefont {Henage}}, \bibinfo {author} {\bibfnamefont
  {T.~A.}\ \bibnamefont {Johnson}}, \bibinfo {author} {\bibfnamefont {T.~G.}\
  \bibnamefont {Walker}}, \ and\ \bibinfo {author} {\bibfnamefont
  {M.}~\bibnamefont {Saffman}},\ }\href@noop {} {\bibfield  {journal} {\bibinfo
   {journal} {Phys. Rev. Lett.}\ }\textbf {\bibinfo {volume} {104}},\ \bibinfo
  {pages} {010503} (\bibinfo {year} {2010})}\BibitemShut {NoStop}%
\bibitem [{\citenamefont {Zhang}\ \emph {et~al.}(2010)\citenamefont {Zhang},
  \citenamefont {Isenhower}, \citenamefont {Gill}, \citenamefont {Walker},\
  and\ \citenamefont {Saffman}}]{Zhang2010}%
  \BibitemOpen
  \bibfield  {author} {\bibinfo {author} {\bibfnamefont {X.~L.}\ \bibnamefont
  {Zhang}}, \bibinfo {author} {\bibfnamefont {L.}~\bibnamefont {Isenhower}},
  \bibinfo {author} {\bibfnamefont {A.~T.}\ \bibnamefont {Gill}}, \bibinfo
  {author} {\bibfnamefont {T.~G.}\ \bibnamefont {Walker}}, \ and\ \bibinfo
  {author} {\bibfnamefont {M.}~\bibnamefont {Saffman}},\ }\href@noop {}
  {\bibfield  {journal} {\bibinfo  {journal} {Phys. Rev. A}\ }\textbf {\bibinfo
  {volume} {82}},\ \bibinfo {pages} {030306(R)} (\bibinfo {year}
  {2010})}\BibitemShut {NoStop}%
\bibitem [{\citenamefont {Zhang}\ \emph
  {et~al.}(2012{\natexlab{b}})\citenamefont {Zhang}, \citenamefont {Gill},
  \citenamefont {Isenhower}, \citenamefont {Walker},\ and\ \citenamefont
  {Saffman}}]{XZhang2012}%
  \BibitemOpen
  \bibfield  {author} {\bibinfo {author} {\bibfnamefont {X.~L.}\ \bibnamefont
  {Zhang}}, \bibinfo {author} {\bibfnamefont {A.~T.}\ \bibnamefont {Gill}},
  \bibinfo {author} {\bibfnamefont {L.}~\bibnamefont {Isenhower}}, \bibinfo
  {author} {\bibfnamefont {T.~G.}\ \bibnamefont {Walker}}, \ and\ \bibinfo
  {author} {\bibfnamefont {M.}~\bibnamefont {Saffman}},\ }\href@noop {}
  {\bibfield  {journal} {\bibinfo  {journal} {Phys. Rev. A}\ }\textbf {\bibinfo
  {volume} {85}},\ \bibinfo {pages} {042310} (\bibinfo {year}
  {2012}{\natexlab{b}})}\BibitemShut {NoStop}%
\bibitem [{\citenamefont {Walker}\ and\ \citenamefont
  {Saffman}(2008)}]{Walker2008}%
  \BibitemOpen
  \bibfield  {author} {\bibinfo {author} {\bibfnamefont {T.~G.}\ \bibnamefont
  {Walker}}\ and\ \bibinfo {author} {\bibfnamefont {M.}~\bibnamefont
  {Saffman}},\ }\href@noop {} {\bibfield  {journal} {\bibinfo  {journal} {Phys.
  Rev. A}\ }\textbf {\bibinfo {volume} {77}},\ \bibinfo {pages} {032723}
  (\bibinfo {year} {2008})}\BibitemShut {NoStop}%
\bibitem [{\citenamefont {Brion}\ \emph {et~al.}(2007)\citenamefont {Brion},
  \citenamefont {Mouritzen},\ and\ \citenamefont {M\o{}lmer}}]{Brion2007c}%
  \BibitemOpen
  \bibfield  {author} {\bibinfo {author} {\bibfnamefont {E.}~\bibnamefont
  {Brion}}, \bibinfo {author} {\bibfnamefont {A.~S.}\ \bibnamefont
  {Mouritzen}}, \ and\ \bibinfo {author} {\bibfnamefont {K.}~\bibnamefont
  {M\o{}lmer}},\ }\href@noop {} {\bibfield  {journal} {\bibinfo  {journal}
  {Phys. Rev. A}\ }\textbf {\bibinfo {volume} {76}},\ \bibinfo {pages} {022334}
  (\bibinfo {year} {2007})}\BibitemShut {NoStop}%
\bibitem [{\citenamefont {Saffman}\ and\ \citenamefont
  {M\o{}lmer}(2009)}]{Saffman2009b}%
  \BibitemOpen
  \bibfield  {author} {\bibinfo {author} {\bibfnamefont {M.}~\bibnamefont
  {Saffman}}\ and\ \bibinfo {author} {\bibfnamefont {K.}~\bibnamefont
  {M\o{}lmer}},\ }\href@noop {} {\bibfield  {journal} {\bibinfo  {journal}
  {Phys. Rev. Lett.}\ }\textbf {\bibinfo {volume} {102}},\ \bibinfo {pages}
  {240502} (\bibinfo {year} {2009})}\BibitemShut {NoStop}%
\bibitem [{\citenamefont {Isenhower}\ \emph {et~al.}(2011)\citenamefont
  {Isenhower}, \citenamefont {Saffman},\ and\ \citenamefont
  {M\o{}lmer}}]{Isenhower2011}%
  \BibitemOpen
  \bibfield  {author} {\bibinfo {author} {\bibfnamefont {L.}~\bibnamefont
  {Isenhower}}, \bibinfo {author} {\bibfnamefont {M.}~\bibnamefont {Saffman}},
  \ and\ \bibinfo {author} {\bibfnamefont {K.}~\bibnamefont {M\o{}lmer}},\
  }\href@noop {} {\bibfield  {journal} {\bibinfo  {journal} {Quant. Inf.
  Proc.}\ }\textbf {\bibinfo {volume} {10}},\ \bibinfo {pages} {755} (\bibinfo
  {year} {2011})}\BibitemShut {NoStop}%
\bibitem [{\citenamefont {Gottesman}\ and\ \citenamefont
  {Chuang}(1999)}]{Gottesman1999}%
  \BibitemOpen
  \bibfield  {author} {\bibinfo {author} {\bibfnamefont {D.}~\bibnamefont
  {Gottesman}}\ and\ \bibinfo {author} {\bibfnamefont {I.~L.}\ \bibnamefont
  {Chuang}},\ }\href@noop {} {\bibfield  {journal} {\bibinfo  {journal}
  {Nature}\ }\textbf {\bibinfo {volume} {402}},\ \bibinfo {pages} {390}
  (\bibinfo {year} {1999})}\BibitemShut {NoStop}%
\bibitem [{\citenamefont {Zhang}\ \emph {et~al.}(2011)\citenamefont {Zhang},
  \citenamefont {Robicheaux},\ and\ \citenamefont {Saffman}}]{SZhang2011}%
  \BibitemOpen
  \bibfield  {author} {\bibinfo {author} {\bibfnamefont {S.}~\bibnamefont
  {Zhang}}, \bibinfo {author} {\bibfnamefont {F.}~\bibnamefont {Robicheaux}}, \
  and\ \bibinfo {author} {\bibfnamefont {M.}~\bibnamefont {Saffman}},\
  }\href@noop {} {\bibfield  {journal} {\bibinfo  {journal} {Phys. Rev. A}\
  }\textbf {\bibinfo {volume} {84}},\ \bibinfo {pages} {043408} (\bibinfo
  {year} {2011})}\BibitemShut {NoStop}%
\bibitem [{\citenamefont {Barahona}(1982)}]{Barahona1982}%
  \BibitemOpen
  \bibfield  {author} {\bibinfo {author} {\bibfnamefont {F.}~\bibnamefont
  {Barahona}},\ }\href@noop {} {\bibfield  {journal} {\bibinfo  {journal} {J.
  Phys. A}\ }\textbf {\bibinfo {volume} {15}},\ \bibinfo {pages} {3241}
  (\bibinfo {year} {1982})}\BibitemShut {NoStop}%
\bibitem [{\citenamefont {Dalibard}\ \emph {et~al.}(1992)\citenamefont
  {Dalibard}, \citenamefont {Castin},\ and\ \citenamefont
  {M\o{}lmer}}]{Dalibard1992}%
  \BibitemOpen
  \bibfield  {author} {\bibinfo {author} {\bibfnamefont {J.}~\bibnamefont
  {Dalibard}}, \bibinfo {author} {\bibfnamefont {Y.}~\bibnamefont {Castin}}, \
  and\ \bibinfo {author} {\bibfnamefont {K.}~\bibnamefont {M\o{}lmer}},\
  }\href@noop {} {\bibfield  {journal} {\bibinfo  {journal} {Phys. Rev. Lett.}\
  }\textbf {\bibinfo {volume} {68}},\ \bibinfo {pages} {580} (\bibinfo {year}
  {1992})}\BibitemShut {NoStop}%
\bibitem [{\citenamefont {Rao}\ and\ \citenamefont
  {M\o{}lmer}(2013)}]{Rao2013}%
  \BibitemOpen
  \bibfield  {author} {\bibinfo {author} {\bibfnamefont {D.~D.~B.}\
  \bibnamefont {Rao}}\ and\ \bibinfo {author} {\bibfnamefont {K.}~\bibnamefont
  {M\o{}lmer}},\ }\href@noop {} {\bibfield  {journal} {\bibinfo  {journal}
  {arXiv:1304.4466}\ } (\bibinfo {year} {2013})}\BibitemShut {NoStop}%
\end{thebibliography}

%

\end{document}